\documentclass{aa}
\usepackage{graphicx}
\usepackage[varg]{txfonts}
\usepackage{natbib}
\bibpunct{(}{)}{;}{a}{}{,}
\usepackage{array}
\usepackage{xspace}
\usepackage{lscape}
\usepackage{url}

\usepackage[hyperfootnotes=false, linktocpage=true, breaklinks=true, colorlinks=true, linkcolor=blue, citecolor=blue, urlcolor=blue]{hyperref}
\usepackage[all]{hypcap}

\usepackage{mathtools}
\usepackage[figuresright]{rotating}

\newcommand{\cv}{\ion{C}{v}\xspace}
\newcommand{\cvi}{\ion{C}{vi}\xspace}
\newcommand{\nvii}{\ion{N}{vii}\xspace}

\newcommand{\ovii}{\ion{O}{vii}\xspace}
\newcommand{\oviii}{\ion{O}{viii}\xspace}

\begin{document} 

\title{Plasma code for astrophysical charge exchange emission at X-ray wavelengths}

\author {Liyi Gu \inst{1} 
\and 
Jelle Kaastra \inst{1,2} 
\and 
A.J.J. Raassen\inst{1,3} }

\institute{
SRON Netherlands Institute for Space Research, Sorbonnelaan 2, 3584 CA Utrecht, The Netherlands \\ \email{L.Gu@sron.nl}
\and 
Leiden Observatory, Leiden University, PO Box 9513, 2300 RA Leiden, The Netherlands 
\and 
Astronomical Institute ``Anton Pannekoek'', Science Park 904, 1098 XH Amsterdam, University of Amsterdam, The Netherlands }
%\institute{SRON Netherlands Institute for Space Research, Sorbonnelaan 2, 3584 CA Utrecht, The Netherlands}

%\date{Received 15 April 2015 / Accepted 22 May 2015}

%==============%
\abstract
%==============%
{ 
Charge exchange X-ray emission provides unique insights into the interactions between cold and hot astrophysical plasmas.
Besides its own profound science, this emission is also technically crucial to all observations in the X-ray band, since charge exchange with 
the solar wind often contributes a significant foreground component that contaminates the signal of interest. By approximating the cross
sections resolved to $n$ and $l$ atomic subshells, and carrying out complete radiative cascade calculation, we create a 
new spectral code to evaluate the charge exchange emission in the X-ray band. Comparing to collisional thermal emission, charge 
exchange radiation exhibits enhanced lines from large-$n$ shells to the ground, as well as large forbidden-to-resonance ratios of 
triplet transitions. Our new model successfully reproduces an observed high-quality spectrum of comet C/2000 WM1 (LINEAR), which emits purely 
by charge exchange between solar wind ions and cometary neutrals. It demonstrates that a proper charge exchange model will
allow us to probe remotely the ion properties, including charge state, dynamics, and composition, at the interface between
the cold and hot plasmas.
}


\authorrunning{L. Gu et al.}
\titlerunning{New Charge Exchange Model}

\maketitle

%====================%
\section{Introduction}
%====================%

Charge exchange (CX hereafter) occurs when an atom collides with a multi-charged ion. It produces X-ray
line emission if the ion charge is reasonably large. The first discovery of CX X-ray emission
in astronomical objects was made by observing the comet C/Hyakutake 1996 B2 with $ROSAT$ telescope (Lisse et al. 1996;
Cravens 1997). This opened up a new research field to X-ray astronomers, and the related study soon expanded to 
various types of celestial objects. The CX process can partly explain the soft X-ray ($< 2$ keV) spectrum of
the polar component in Jupiter and other planets in the solar system (e.g., Bhardwaj 2006; Branduardi-Raymont et al. 2007). As suggested
in many papers, e.g., Cox (1998), Cravens (2000), Snowden et al. (2004), and Fujimoto et al. (2007), the geocoronal
and heliospheric area is also glowing in soft X-rays by the CX between solar wind ions and interstellar neutral atoms.  
CX is also a potential mechanism for X-ray emission from the North Polar Spur region (Lallement 2009), stellar winds
of supergiants (Pollock 2007), shock rims of supernova remnants (Katsuda et al. 2011; Cumbee et al. 2014), starburst 
galaxies (Tsuru et al. 2007; Liu et al. 2011), and even clusters of galaxies (Walker et al. 2015).

The astronomical discoveries of CX emission have also stimulated theoretical and laboratory studies of this process.
Recent atomic calculations based on the quantum mechanical close coupling method (e.g., Wu et al. 2011; Nolte et al. 2012)
provided so far the most reliable estimate of the cross section for single electron capture in the low-velocity regime.
Laboratory investigations broadened the species of neutral reactants, from the hydrogen atom to various molecules (e.g., 
$\rm CO$, $\rm H_{2}O$), to match better with the real conditions, especially for comets (Bodewits et al. 2006; Wargelin et al. 2008).

To link the atomic and astrophysical work on the CX process, we present in this paper a new spectral model featuring
the most recent atomic data and a complete radiative transition calculation. Another model for CX was introduced by Smith
et al. (2014), which is more focused on the transition calculation rather than the actual reaction cross section. In \S2, we
describe the physical assumption of our model. In \S3, we show in detail how the CX line emission is calculated 
from basic atomic data. In \S4, we demonstrate our model by applying it to real data.

%====================%
\section{Assumptions}
%====================%

Our CX emission model is calculated based on three key assumptions. First, the charge transfer rates are obtained based 
on single electron capture in ion-neutral collision. Although multi-electron neutral targets might be important 
for some environments (e.g., comets), where the channels of multi-electron capture do exist (e.g., Bodewits et al. 
2006), the data available on such reactions are much less complete than those for single electron capture. As shown
in the experimental results of e.g., Greenwood et al. (2001), single capture can provide a first-order 
approximation to multi-electron reactions. Secondly, as a related issue, we assume
that the CX with atomic hydrogen is a reasonable representative to the real case, in terms of the cross section dependences on  
ion velocity and captured electron state (see details in \S3). To correct for the helium atom, we approximate the 
helium cross section using the scaling rule of Janev \& Gallagher (1984),
\begin{equation}
\frac{\sigma_{\rm He}} {\sigma_{\rm H}} = \frac{N_{\rm He}} {N_{\rm H}} \left(\frac{I_{\rm H}}{I_{\rm He}}\right)^{2},
\end{equation}       
where $N_{\rm He}$ and $N_{\rm H}$ are the numbers of valence electrons, and $I_{\rm He}$ and $I_{\rm H}$ are the ionization
potentials. For plasmas with cosmological abundances (10\% and 90\% in number for helium and hydrogen, respectively), 
the combined CX cross section can be derived from the pure hydrogen value by $\sigma = 0.96 \sigma_{\rm H}$.

As the third assumption, the radiative processes related
to free electrons, e.g., collisional excitation and radiative/dielectron recombination, are ignored in the modeling.
It will prevent the CX-induced transitions, which usually involves large-$n$ shells (\S3.2), from being overpowered
in emission by the collisional excitation dominating the small-$n$ shells. This assumption can be validated because the ionic CX has much larger cross section 
than the electronic processes at X-ray energies.

%====================%
\section{Method}
%====================%

To calculate CX line emission, we first determine the ion state population after
electron capture, and then solve the possible radiative cascading pathways to 
the ground state. The first step can be further divided into three components, i.e., the 
total capture cross sections, the cross sections into each $n-$, and $l-$ resolved
level. 

The main challenge is that the current atomic data for $n-$ and $l-$ resolved
cross sections are far from complete. It is hence necessary to investigate the available
data for intrinsic scaling relations among $n-$ and $l-$ resolved parameters, as described in \S3.2 and \S3.3, respectively.
In practice, we collected all available cross sections from literature, and applied the
derived scaling relations when the actual data are absent.

%====================%
\subsection{Total cross sections}
%====================%
%============================
%  TABLE: Data sources
%
\begin{table*}[!]
\begin{minipage}[t]{\hsize}
\setlength{\extrarowheight}{3pt}
\caption{Collected charge exchange data}
\label{lines_table}
\centering
\small
\renewcommand{\footnoterule}{}
\begin{tabular}{l c c c || l c c c}
\hline \hline
Ion &  method\tablefootmark{a}  & type\tablefootmark{b} & reference &  Ion &  method\tablefootmark{a}  & type\tablefootmark{b} & reference \\
\hline
$\rm Be^{4+}$   & UDWA & $nl$ & Ryufuku (1982)      & $\rm N^{7+}$  & MOCC & $nl$ & Harel et al. (1998) \\
$\rm B^{5+}$  & UDWA & $nl$ & Ryufuku (1982)        & $\rm O^{8+}$  & MOCC & $nl$ & Harel et al. (1998) \\
$\rm C^{6+}$  & UDWA & $nl$ & Ryufuku (1982)        & $\rm Ar^{q+}$ ($15 \leq q \leq 18$)  & CTMC & $nl$ & Whyte et al. (1998)\\
$\rm O^{8+}$  & UDWA & $nl$ & Ryufuku (1982)        & $\rm Ne^{10+}$  & CTMC & $n$ & Perez et al. (2001)\\
$\rm O^{8+}$  & MOCC & $nl$ & Shipsey et al. (1983) & $\rm Ar^{18+}$  & CTMC & $n$ & Perez et al. (2001)\\
$\rm Be^{4+}$  & AOCC & $nl$ & Fritsch \& Lin (1984) & $\rm Fe^{18+}$  & CTMC & $n$ & Perez et al. (2001)\\
$\rm B^{5+}$  & AOCC & $nl$ & Fritsch \& Lin (1984) & $\rm O^{q+}$ ($5 \leq q \leq 8$) & comp. & $nl$ & Rakovi$\rm \acute{c}$ et al. (2001)\\
$\rm C^{4+}$  & AOCC & $nl$ & Fritsch \& Lin (1984) & $\rm Li^{3+}$  & CTMC & $nl$ & Errea et al. (2004)\\
$\rm C^{6+}$  & AOCC & $nl$ & Fritsch \& Lin (1984) & $\rm Ne^{10+}$  & CTMC & $nl$ & Errea et al. (2004)\\
$\rm O^{8+}$  & AOCC & $nl$ & Fritsch \& Lin (1984) & $\rm Ar^{q+}$ ($15 \leq q \leq 18$)  & CTMC & $nl$ & Schultz et al. (2010)\\
$\rm Fe^{q+}$ ($5 \leq q \leq 26$)  & comp. & total & Phaneuf et al. (1987) & $\rm N^{6+}$ & QMOCC & $nlS$ & Wu et al. (2011)\\
$\rm C^{q+}$ ($1 \leq q \leq 6$)  & comp. & total   & Janev et al. (1988) & $\rm O^{6+}$ & QMOCC & $nlS$ & Wu et al. (2012)\\
$\rm O^{q+}$ ($1 \leq q \leq 8$)  & comp. & total   & Janev et al. (1988) & $\rm C^{5+}$ & QMOCC & $nlS$ & Nolte et al. (2012)\\
$\rm C^{6+}$  & comp. & $nl$ & Janev et al. (1993)  & $\rm Fe^{q+}$ ($25 \leq q \leq 26$) & MCLZ & $n$ & Mullen et al. (2015)\\
$\rm O^{8+}$  & comp. & $nl$ & Janev et al. (1993)  & $\rm He^{2+}$ & EXP & total & Fite et al. (1962)\\
$\rm C^{6+}$  & CTMC & $nl$ & Olson \& Schultz (1989) & $\rm He^{q+}$ ($1 \leq q \leq 2$) & EXP & total & Olson et al. (1977)\\
$\rm O^{8+}$  & CTMC & $nl$ & Olson \& Schultz (1989) & $\rm O^{q+}$ ($3 \leq q \leq 8$) & EXP & total & Meyer et al. (1979)\\
$\rm Li^{3+}$  & CDWA & $n$ & Belki{\'c} (1991)     & $\rm Si^{9+}$ & EXP & total & Meyer et al. (1979)\\
$\rm Be^{4+}$  & CDWA & $n$ & Belki{\'c} (1991)     & $\rm Fe^{15+}$ & EXP & total & Meyer et al. (1979)\\
$\rm B^{5+}$  & CDWA & $n$ & Belki{\'c} (1991)      & $\rm C^{6+}$ & EXP & total & Meyer et al. (1985)\\
$\rm C^{6+}$  & CDWA & $nl$ & Belki{\'c} (1991)     & $\rm N^{q+}$ ($6 \leq q \leq 7$) & EXP & total & Meyer et al. (1985)\\
$\rm N^{7+}$  & CDWA & $n$ & Belki{\'c} (1991)      & $\rm O^{q+}$ ($7 \leq q \leq 8$) & EXP & total & Meyer et al. (1985)\\
$\rm O^{8+}$  & CDWA & $nl$ & Belki{\'c} (1991)     & $\rm F^{q+}$ ($8 \leq q \leq 9$) & EXP & total & Meyer et al. (1985)\\
$\rm He^{2+}$  & MOCC & $nl$ & Harel et al. (1998)  & $\rm Ne^{q+}$ ($9 \leq q \leq 10$) & EXP & total & Meyer et al. (1985)\\
$\rm Li^{3+}$  & MOCC & $nl$ & Harel et al. (1998)  & $\rm B^{q+}$ ($3 \leq q \leq 5$) & EXP & total & Crandall et al. (1979)\\
$\rm Be^{4+}$  & MOCC & $nl$ & Harel et al. (1998)  & $\rm C^{q+}$ ($3 \leq q \leq 4$) & EXP & total & Crandall et al. (1979)\\
$\rm B^{5+}$  & MOCC & $nl$ & Harel et al. (1998)   & $\rm N^{q+}$ ($3 \leq q \leq 4$) & EXP & total & Crandall et al. (1979)\\
$\rm C^{6+}$  & MOCC & $nl$ & Harel et al. (1998)   & $\rm O^{q+}$ ($5 \leq q \leq 6$) & EXP & total & Crandall et al. (1979)\\
\hline
\hline

\end{tabular}
\end{minipage}
 \tablefoot{
   \tablefoottext{a}{Methods include: UDWA (unitarized distorted-wave approximation), MOCC (molecular-orbital close-coupling),
     AOCC (atomic-orbital close-coupling), comp. (data compilation), CTMC (classical trajectory Monte Carlo), CDWA
     (continuum distorted-wave approximation), QMOCC (quantum molecular-orbital close-coupling), MCLZ (multichannel Landau-Zener),
     and EXP (experiment);}
   \tablefoottext{b}{Data types are: total (total cross section only), $n$ (principle quantum number $n$-resolved cross section),
     $nl$ ($nl$-resolved), and $nlS$ ($nlS$-resolved).}
 }

\end{table*}

%============================
%  FIG: Total cross sections
%
\begin{figure*}[!]
\centering
\resizebox{0.5\hsize}{!}{\hspace{-1cm}\includegraphics[angle=0]{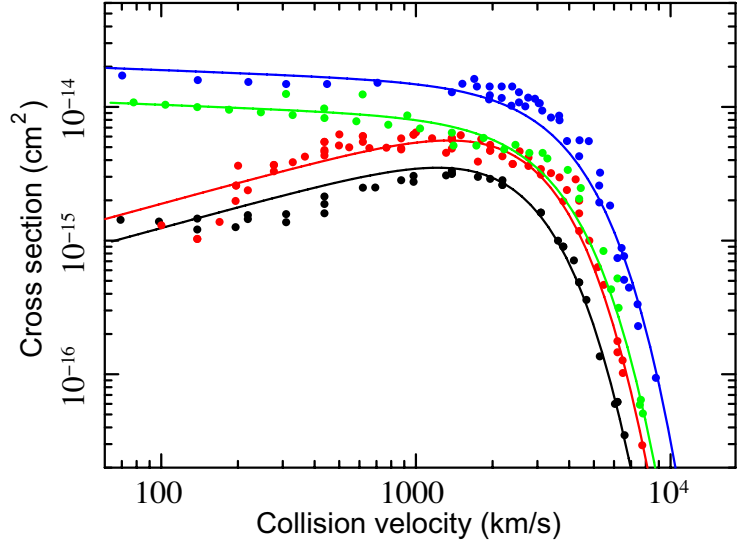}}
\caption{Total cross sections, as functions of collision velocity, for $\rm B^{5+}$ (black), $\rm O^{8+}$ (red), $\rm Ne^{10+}$ (green), and $\rm Ar^{18+}$ (blue) 
interacting with a hydrogen atom. The data points are actual values from previous calculations, and solid lines are the scaling law described in \S3.1.
\vspace{0.5cm}
}
\label{totalcs_fig}
\end{figure*}
%============================

It is commonly believed that CX has a very large cross section
compared to electronic processes, typically of order $10^{-15}-10^{-14}$ cm$^{-2}$.
To obtain the cross section as a function of collision velocity $v$, we compiled 
the results reported in previous theoretical calculations (Ryufuku 1982; Shipsey et al. 1983; Fritsch \& Lin et al. 1984; Phaneuf et al. 1987;
Janev et al. 1988, 1993; Olson \& Schultz 1989; Belki{\'c} 1991; Harel et al. 1998; Whyte et al. 1998; Perez et al. 2001; Errea et al. 2004; Schultz et al. 2010; Wu et al. 2011; Nolte et al. 2012) and 
experimental measurements (Fite et al. 1962; Olson et al. 1977; Meyer et al. 1979, 1985; Crandall et al. 1979). A complete list of the
data sources is shown in Table 1. For the theoretical calculations, we employed a practical criterion reported in Janev et al. (1993), which is dependent on
the calculation method, to assess the energy range of validity. 

As shown in, e.g., Wargelin et al. (2008),
the average cross sections usually exhibit a linear dependence on ion charge $q$. Such a feature is also seen in Figure 1. 
In addition, due to the non-resonant effect (e.g., Janev \& Winter 1985), the cross sections for small-$q$ reactions, e.g., $\rm B^{5+}$ with the H atom, 
exhibit a maximum at certain $v \leq v_{0}$ (where $v_{0}$ is the orbital velocity of bound electrons), and exponential decrease towards low energy.
For large-$q$ ions, e.g., $\rm Ne^{10+}$ and $\rm Ar^{18+}$, such an effect diminishes, as the number of channels for resonant reactions becomes large. In the high energy 
regime ($v > v_{0}$), both small-$q$ and large-$q$ 
reactions fall off, as charge transfer into discrete levels become strongly coupled with the continuum, and the ionization process starts to dominate. 
To combine the $q-$ and $v-$ dependence, we use a scaling law refined from 
Janev \& Smith (1993), 

\begin{equation}
\begin{multlined}
\sigma = a_{1} q E_{s}^{a_{2}} {\rm ln}\left( \frac{a_{3}}{E_{s}} + a_{4} \right) \left( 1 + \frac{E_{s}}{a_{5}} \right)^{a_{6}}, 
\end{multlined}
\end{equation}
 
\noindent where $E_{s} = E / q^{0.43}$ is a scaled version of the collision energy $E$ given in eV/amu, and $a_{1}$ to $a_{6}$ are $q-$ dependent fitting parameters. 
The average best-fit values for $a_{1}$ to $a_{6}$ are (4.6, 1.0, 0.2, 1.0, 83., -8.9)    
and (0.3, 0.1, 1., 10., 158., -9.9), for $3 < q < 10$ and $q \geq 10$, respectively. As shown in Figure 1, the scaling relation in general reproduces well 
the cross sections of all ions. Some residuals are seen at $v \leq 500$ km/s, where the data oscillate around the fitting curve for small-$q$ species. This is probably
caused by the discrete nature of product ion energy levels (e.g., Ryufuku 1980). In practice, since the actual data for such small-$q$ species are usually available 
in the literature, the bias on the scaling relation can barely affect our model. Eq.2 is then used to calculate cross sections for ions that are still missing
in previous publications.

%====================%
\subsection{$n-$shell populations}
%====================%

%============================
%  FIG: n-resolved cross sections
%
\begin{figure*}[!]
\centering
\resizebox{0.6\hsize}{!}{\hspace{-1cm}\includegraphics[angle=0]{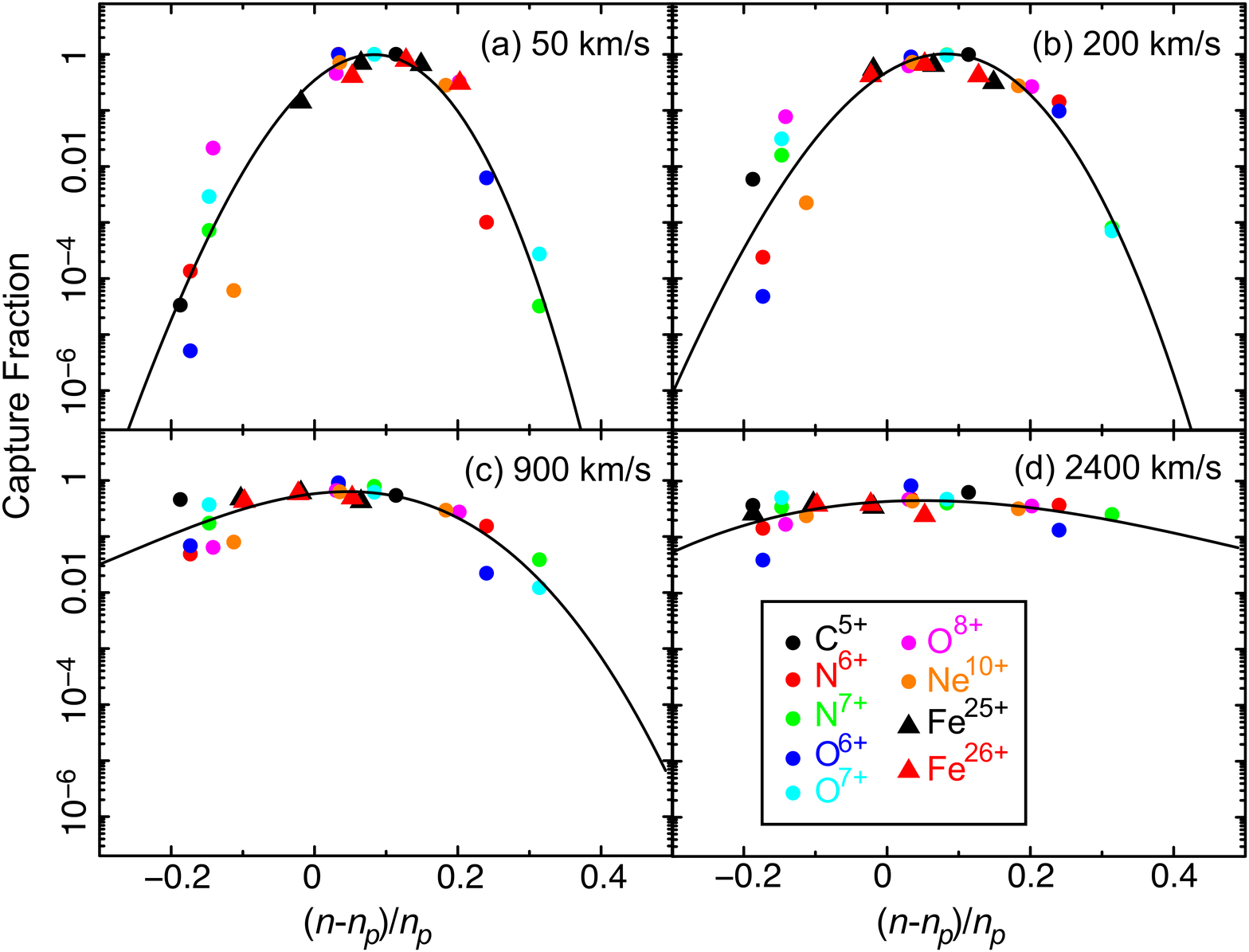}}
\caption{Fractions of single electron capture into $n$ shell, at $v = 50$ km $\rm s^{-1}$ (a), 200 km $\rm s^{-1}$ (b), 900 km $\rm s^{-1}$ (c), and 2400 km $\rm s^{-1}$ (d), as a function of $(n - n_{\rm p})/n_{\rm p}$. Different sets of data points represent different projectile ions, as indicated in (d), and the solid lines are the polynomial fitting to the data.
\vspace{0.5cm}
}
\label{n-cs_fig}
\end{figure*}
%============================

The CX probability reaches its maximum when the two energy states, before and after the transition, are most close to each other. For low-speed
collisions ($v << v_{0}$), the potential energy dominates the interaction, and the principle quantum number $n_{\rm p}$ of the most populated energy level
can be written by
\begin{equation}
n_{\rm p} = \sqrt{\frac{I_{\rm H}}{2 I_{\rm t}}} q \left( 1 + \frac{q-1}{\sqrt{2q}} \right)^{-0.5}
\end{equation}       
(Janev \& Winter 1985), where $I_{\rm H}$ and $I_{\rm t}$ are the ionization potentials of H and the target atom, respectively. This means that the peak level 
is determined by the combined potential of the projectile ion and target atom. For most ions, $n_{\rm p}$ is much larger than unity. In the high-speed regime ($v \sim v_{0}$),
the collision dynamics become more important for the capture process, and the peak population level is gradually smeared out among several adjacent shells.

We compiled the velocity-dependent, $n-$ resolved cross sections for reactions involving Be, B, C, N, O, Ne, and Fe ions from theoretical calculations 
(Ryufuku 1982; Shipsey et al. 1983; Fritsch \& Lin et al. 1984; Belki{\'c} et al. 1992; Janev et al. 1993; Toshima \& Tawara 1995; Harel et al. 1998; 
Rakovi$\rm \acute{c}$ et al. 2001; Errea et al. 2004; Nolte et al. 2012; Wu et al. 2011, 2012; Mullen et al. 2015). As shown in Figure 2, the capture cross sections are normalized 
to the total value of each ion, and plotted against $(n - n_{\rm p}) / n_{\rm p}$. This brings all the ions on a roughly similar $n-$ distribution function,
which reconfirms the result in Janev \& Winter (1985). The velocity dependence of the $n-$ populations also agrees with the consensus described above. 
As seen in Figure 2, for each velocity, we fit the $n-$ distribution by a phenomenological third-degree polynomial curve, which well reproduces the data.
Some minor deviations, typically a few percent of the total cross section, are seen at low-capture shells. The same fitting was done for ten other velocity points
to fully cover the range of $50 \leq v \leq 5000$ km $\rm s^{-1}$. Details of the fitting procedure can be found in Appendix A. As a first-order approximation,
we assumed that all the CX processes with hydrogen atom targets follow the same profile in the $(n - n_{\rm p}) / n_{\rm p}$ versus $\sigma$ space, and share a similar
velocity dependence. In such a way, the $n-$ resolved populations were calculated for all the rest ions. When applying the $n-$ distribution to astrophysical plasmas
contaminated by helium, the actual peak $n$ would be slightly lower than the calculated value, since the helium target has a larger ionization potential than
atomic hydrogen. Assuming the cosmic abundances, this would bring an uncertainties of $\leq 10$\% to the obtained $n-$ resolved cross sections.

%====================%
\subsection{$l-$subshell populations}
%====================%

%============================
%  FIG: l-resolved cross sections
%
\begin{figure*}[!]
\centering
\resizebox{0.6\hsize}{!}{\hspace{-1cm}\includegraphics[angle=0]{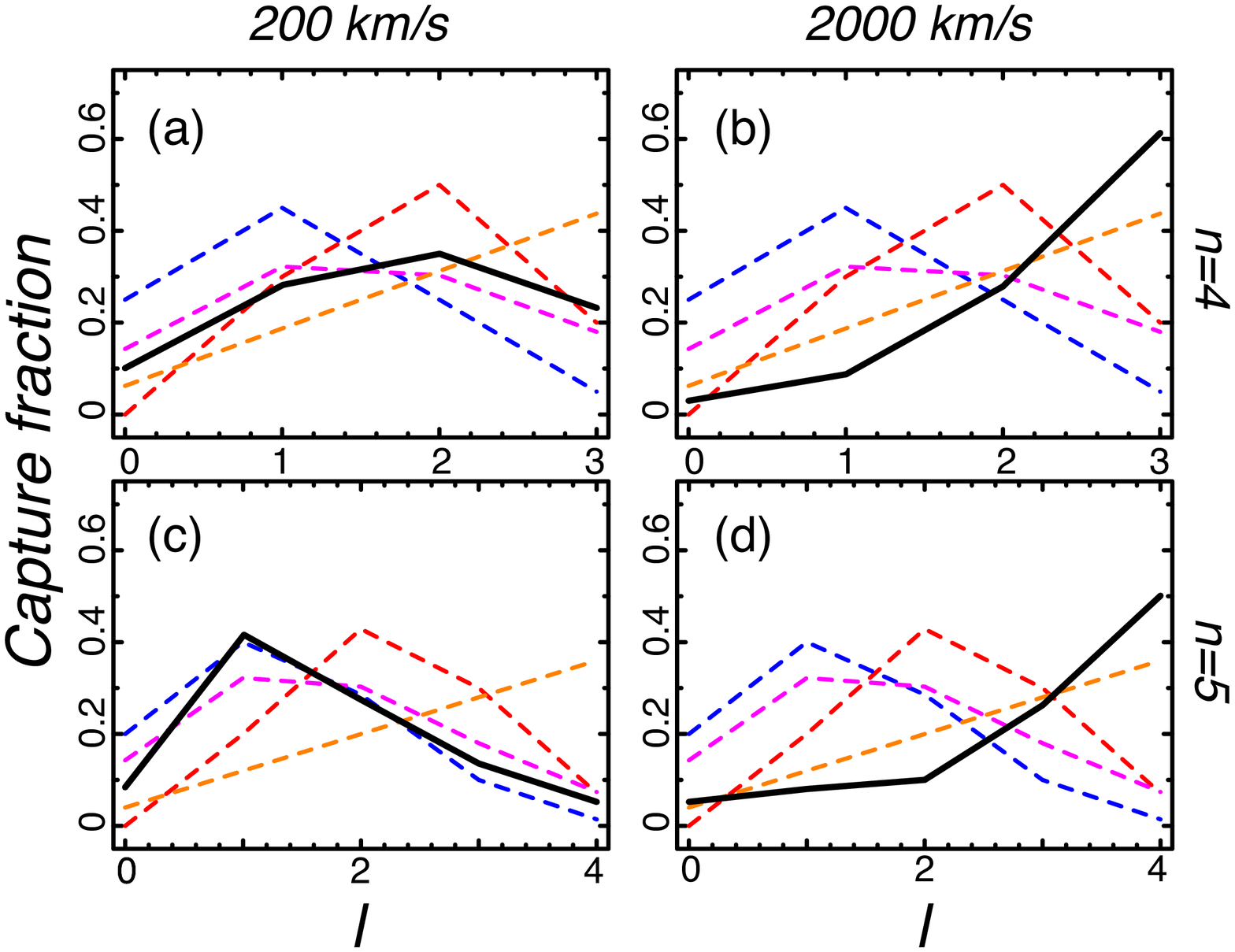}}
\caption{Averaged fractions of $l-$dependent capture for $\rm N^{7+}$ and $\rm O^{7+}$, plotted as a function of $l$. Black solid lines are the data
  from theoretical calculations, and the red, blue, magenta, and orange dashed lines are empirical functions shown in Eqs.4, 5, 6, 8, respectively.
  Panels (a) and (b) are the capture for $n = 4$ at $v = 200$ km $\rm s^{-1}$ and 2000 km $\rm s^{-1}$, respectively, and (c) and (d) are those for
  $n = 5$.
\vspace{0.5cm}
}
\label{l-cs_fig}
\end{figure*}
%============================

Besides the $n-$ dependent capture, it is known that the CX process also exhibits strong selective properties with respect to the final electron orbit angular 
momentum $l$. As discussed in, e.g., Janev \& Winter (1985) and Suraud et al. (1991), the $l-$ selectivity 
is very sensitive to the collision velocity and is often governed by high-order processes in the transition (e.g., rotational mixing). Typically the 
$l$ distribution is approximated as a function of $n$, $l$, and $q$, in at least five forms shown as follows:
\begin{equation}
W^{\rm l1}_{n}(l) = (2l + 1) \frac{[(n-1)!]^{2}} {(n+l)!(n-l-1)!}     \:\:\:\:\:\:\:\:\:\:\:\:\:\:\:\:\:  \bf(Low \: energy \: I),
\end{equation}       
\begin{equation}
W^{\rm l2}_{n}(l) = l(l+1)(2l + 1) \frac{(n-1)!(n-2)!} {(n+l)!(n-l-1)!}     \:\:\:\:     \bf(Low \: energy \: II),
\end{equation}  
 \begin{equation}
W^{\rm se}_{n}(l) = \left(\frac{2l+1}{q}\right) {\rm exp}\left[-\frac{l(l+1)}{q}\right]     \:\:\:\:\:\:\:\:\:\:\:\:\:\:\:\:\:\:\:\:     \bf(Separable),
\end{equation}       
\begin{equation}
W^{\rm ev}_{n}(l) = 1 / n     \:\:\:\:\:\:\:\:\:\:\:\:\:\:\:\:\:\:\:\:\:\:\:\:\:\:\:\:\:\:\:\:\:\:\:\:\:\:\:\:\:\:\:\:\:\:\:\:\:\:\:\:\:\:\:\:\:  \bf(Even),
\end{equation}       
\begin{equation}
W^{\rm st}_{n}(l) = (2l + 1) / n^{2}     \:\:\:\:\:\:\:\:\:\:\:\:\:\:\:\:\:\:\:\:\:\:\:\:\:\:\:\:\:\:\:\:\:\:\:\:\:\:\:\:\:\:\:\:\:  \bf(Statistical).
\end{equation}

In the low velocity regime ($v \ll v_{0}$), generally the transferred electron does not carry sufficient angular momentum to populate large-$l$ subshells. These
electrons form a peak at $l = 1$ or 2 (e.g., Abramov et al. 1978), which can be roughly described by functions $W^{\rm l1}$, $W^{\rm l2}$, and/or $W^{\rm se}$. 
As the collision velocity increases, the $l$ peak is gradually smeared out by rotational mixing of the coupling Stark states, and the subshells are populated
according to the statistical weight, i.e., $W^{\rm st}$.

It is clear that none of the five functions alone can describe the $l-$ populations for all velocities. Previous work further suggested that the choice of weighting function
also depends on the principle quantum number $n$; the $l-$ distributions of $n \leq n_{\rm p}$ subshells are often found different from those with $n > n_{\rm p}$ (e.g., Janev \& Winter 1985).
To elucidate the $v-$ and $n-$ dependence, we plot in Figure 3(a) and 3(c) the averaged $l-$ distributions at $v = 200$ km $\rm s^{-1}$ for
$\rm N^{7+}$ and $\rm O^{7+}$ ions, based on data from theoretical calculations (Belki{\'c} et al. 1992; Toshima \& Tawara 1995; Rakovi$\rm \acute{c}$ et al. 2001).   
The two dominant shells with $n = 4$ and 5 are used to represent the $n \leq n_{\rm p}$ and $n > n_{\rm p}$ groups, respectively. This is valid since the rest shells contribute less
than 0.1\% of the total rate at $v = 200$ km $\rm s^{-1}$. The $\rm N^{7+}$ and $\rm O^{7+}$ ions exhibit a large capture fraction into $l=1$ for $n=5$, while for $n=4$, 
subshells with $l=2$ and 3 are equally or even more populated than $l=1$. By comparing data with the five weighting functions, we found that the $l-$ populations of
the two ions are best approximated by the $W^{\rm l2}$ function at $n = 5$, while the $n=4$ shells resembles more the $W^{\rm se}$ function. As shown in Figure 3(b) and 3(d),
the same calculation was done at $v = 2000$ km $\rm s^{-1}$, where the two distributions become roughly consistent, and match best with the statistical weight $W^{\rm st}$.
By incorporating the data of other available ions, we determined the preferred weighting function dependent on $v$ and $n$, and applied it to the rest ions (see Appendix B
for details).

%====================%
\subsection{Spectral model}
%====================%

The CX emission line is produced when the captured electron relaxes to the valence shell of the ground state of the product ion. To perform the complete cascade calculation, we 
have obtained the energies and transition probabilities for all the atomic levels up to $n=16$, which exceeds the maximum capture state for all ions used in the code (e.g., 
maximum capture $n = 12$ and 13 for Fe$^{25+}$ + H collision). The FAC code for theoretical atomic calculations (Gu 2008) was used as a baseline tool, and the energies 
were calibrated to those derived with another 
atomic structure code (Cowan 1981), as well as the experimental measurements at the National Institute of Standards and Technology.\footnote{http://www.nist.gov/pml/data/asd.cfm} 
The line spectrum was then calculated from the cascade network given the source term of the CX rates. More features of the new CX model are described in Appendix C.

As described in \S3.2, the CX process can populate the large-$n$ shells, hence line emission from such shells is strongly enhanced compared to collisional thermal radiations.
For instance, the \oviii Ly$\rm \delta$ line at 14.8 $\rm \AA$ is stronger than the Ly$\rm \gamma$ line at 15.2 $\rm \AA$. Similar conditions can be found in many other transitions, e.g., 
the \ovii He$\rm \delta$ line at 17.4 $\rm \AA$, the \nvii Ly$\rm \delta$ line at 19.4 $\rm \AA$, and \cvi Ly$\rm \delta$ line at 26.4 $\rm \AA$. The derived CX spectrum also features a large G 
ratio, i.e., forbidden plus intercombination lines to resonance line ratio, since the collisional excitation process becomes negligible in CX plasmas (\S2).

%====================%
\subsection{Bias from systematic weight}
%====================%

%============================
%  FIG: S-resolved cross sections
%
\begin{figure*}[!]
\centering
\resizebox{0.4\hsize}{!}{\hspace{-1cm}\includegraphics[angle=0]{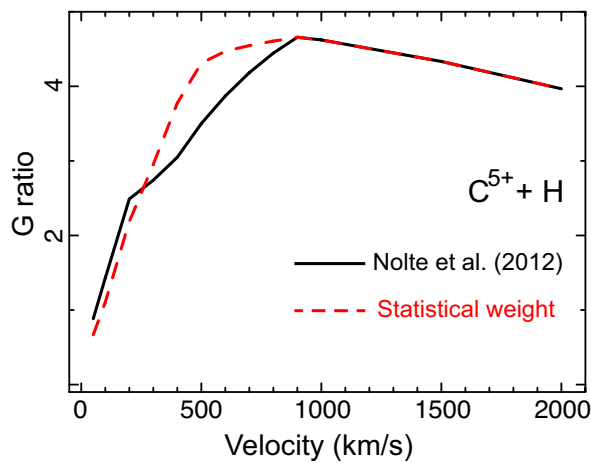}}
\caption{G ratio of \cv He$\alpha$ triplets, calculated based on the triplet-to-singlet ratios reported in Nolte et al. (2012, black), and the value given by statistical weight (red).
\vspace{0.5cm}
}
\label{S-cs_fig}
\end{figure*}
%============================

As reported recently by Nolte et al. (2012), the singlet-to-triplet ratio of the $\rm C^{4+}$ ion, produced from a low-velocity $\rm C^{5+}$ $+ \rm H$ reaction, covers a broad 
range of $0.01 - 100$ for different $n$, $l$, and $v$. The apparent bias from the commonly adopted, statistical value 3 is probably caused by electron-electron 
interaction during the capture. A similar effect was also reported in other papers (e.g., Wu et al. 2011 for $\rm N^{6+}$ $+ \rm H$ reaction), indicating that it could be
a common property. Although the current data are yet too sparse to fully 
implement the $S-$ dependence, it is vital to estimate the induced biases on emission line ratios. As shown in Figure 4, we compare the G ratio calculated based on the data from Nolte et al. (2012), 
with that assuming statistical weight for the $\rm C^{5+}$ $+ \rm H$ reaction. The largest bias is 
seen at $v \sim 500$ km $\rm s^{-1}$,
where the statistical weight is underestimated by about 30\%. The two ratios become roughly consistent at low and high velocity ends. The figure suggests that the current code
can well cover the actual G ratio range, although the derived velocity would have a systematic error up to $200-300$ km $\rm s^{-1}$.

%====================%
\section{Real data fitting}
%====================%
 
To verify the CX model, we fit it to the real data of comet C/2000 WM1 (LINEAR) observed 
by the {\it XMM-Newton} Reflection Grating Spectrometer (RGS), which has a high spectral 
resolution $\sim 0.07$ $\rm \AA$ in soft X-rays (i.e., $5-38$ $\rm \AA$). Comets are the favorite target
for our purpose, since they emit bright X-rays that are exclusively produced by the CX between the highly charged heavy ions in the
solar wind and cometary atmosphere (Dennerl 2010). As a caveat, the current model is based on atomic hydrogen target, 
while the cometary neutrals are mainly moleculars, such as $\rm H_{2}O$ and CO. As reported in
Bodewits et al. (2007) and Mullen et al. (2015), the cross sections of moleculars can be 
roughly approximated by that of H at intermediate and high velocities, while at low velocity (i.e.,
$v \lesssim 100$ km $\rm s^{-1}$), they become different by order of magnitudes. The cometary
CX model with molecular targets will be reported in a following paper (Mullen et al. 2016).

%====================%
\subsection{Observation}
%====================%
%============================
%  FIG: C/2000 wm1
%
\begin{figure*}[!]
\centering
\resizebox{0.5\hsize}{!}{\hspace{-1cm}\includegraphics[angle=0]{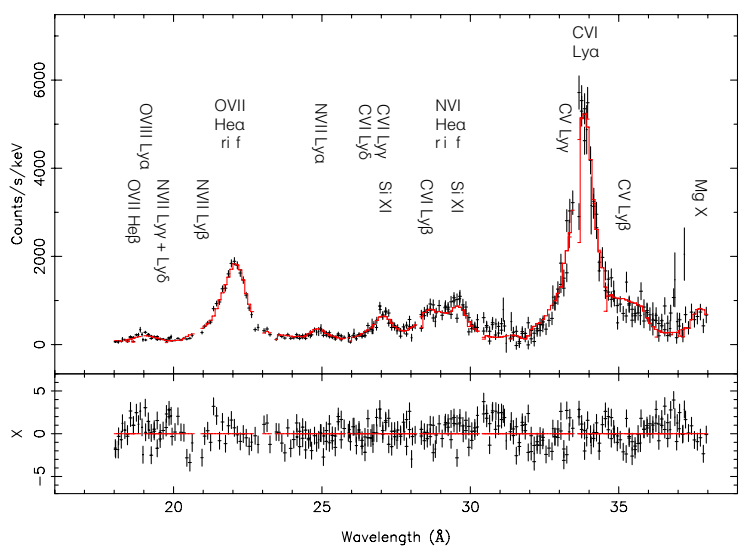}}
\caption{CX fitting (red) to the RGS spectrum (black) of comet C/2000 WM1 (LINEAR).
\vspace{0.5cm}
}
\label{c2000_fig}
\end{figure*}
%============================

Comet C/2000 WM1 was observed by {\it XMM-Newton} on December 2001, for a continuous exposure of about 62 ks. 
The RGS event files were created by using SAS 14.0 and the most recent calibration files. To remove events
outside the field of view along the cross-dispersion direction ($5^{\prime}$), we used only the 18 ks exposure
when the comet was close to the center point of the detector. The background component was approximated 
by the model background spectrum calculated by SAS tool {\tt rgsproc}. To correct the spectral broadening
due to spatial extent of the cometary X-ray halo, we used a $7-30$ $\rm \AA$ image which was observed at the same period with
the MOS1 detector of the European Photon Imaging
Cameras (EPIC) onboard {\it XMM-Newton}. The final RGS data has more than
18000 counts recorded with high spectral resolution, making it so far the best X-ray data for the cometary CX study.

%====================%
\subsection{Results}
%====================%

Our study focuses on the $22 - 38$ $\rm \AA$ ($0.33-0.56$ keV) band of the RGS spectrum. In the fits,
we assume that all the solar wind ions had a same ionization temperature, and collided with
the comet atmosphere at a constant velocity during the exposure. The temperature and velocity are allowed
to vary freely. We can also measure the C, N, O, Mg, and Si abundances of the wind; other elements are not visible
in the RGS energy band. Following the recipe of von Steiger et al. (2000), 
the solar abundances are set to the photospheric values from Grevesse \& Sauval (1998). The CX emissivity,
which is determined by combining the ion and neutral densities, was also set free, although it would degenerate to some extent with 
the metal abundances. As shown in Figure 5,
the RGS spectrum can be fit reasonably well by one CX component, characterized by an ionization temperature of $0.14 \pm 0.01$ keV
and a collision velocity of $200 \pm^{85}_{45}$ km $\rm s^{-1}$. The fitting C-statistic is 401 for a degrees of freedom of 288. The collision velocity is measured primarily by the strong \cvi lines
from different $n$ shells, including the Ly$\rm \alpha$ line at $33.7 \rm \AA$, the Ly$\rm \beta$ line at $28.5 \rm \AA$,
the Ly$\rm \gamma$ line at $27.0 \rm \AA$, and the Ly$\rm \delta$ line at $26.4 \rm \AA$. The \nvii and \ovii
series also assist to the velocity measurement. The smaller lower velocity error compared to the upper one is probably due to the
fact that the line ratios become more sensitive to the velocities at lower collision energies. The best-fit abundance
ratios relative to O, i.e., C/O, N/O, Mg/O, and Si/O, are measured to be $1.9 \pm 0.3$, $1.6 \pm 0.6$, $5 \pm 4$, and $3 \pm 2$, respectively.
These values are roughly consistent with the average solar wind abundances reported in von Steiger et al. (2000).

The derived CX component appears to resemble the slow-type solar wind. It is well known that the slow wind is launched
with a typical ionization state of $0.12-0.14$ keV, which is quite different from that of fast wind (0.07 keV, Feldman et al. 2005). The ionization 
temperature remains nearly the same
in the wind propagation, since the ionization/recombination timescales are often much longer than the travel time.
On the other hand, according to the records of Advanced Composition Explorer (ACE), the solar wind ion speed at the Earth Lagrangian point L1 is
in the range of $250-350$ km $\rm s^{-1}$, in the period of the C/2000 WM1 observation by {\it XMM-Newton}. Since the heliocentric distance 
of the comet was nearly 1 AU, the ion speed at its environment should be close to the ACE record (Neugebauer et al. 2000). Consider that
the comet had a velocity of $\sim 50$ km $\rm s^{-1}$ relative to the Sun, the best-fit collision velocity measured with our CX model appears to
be slightly lower than the wind speed in the comet restframe. This agrees with the picture presented in, e.g., Bodewits et al. (2007), in which
the solar wind is somewhat decelerated in the comet bow shock region.

%====================%
\section{Summary}
%====================%

We developed a new plasma code to calculate charge exchange emission in X-ray band. To overcome the incompleteness in atomic data
of cross sections, we derived scaling laws to approximate the $n-$ and $l-$ distributions for various collision velocities in the range 
of $50-5000$ km $\rm s^{-1}$. The radiative cascading calculation shows characteristic charge exchange emission features, including high-shell 
transition lines as well as large
G ratios of triplets. Our CX model successfully reproduces an observed high resolution X-ray spectrum from comet C/2000 WM1 with reasonable ionization temperature
and collision velocity of the solar wind ions.

\begin{acknowledgements}

SRON is supported financially by NWO, the Netherlands Organization for Scientific Research. 

\end{acknowledgements}

\appendix
%====================%
\section{Fitting the $n-$distributions}
%====================%

%============================
%  TABLE: n-distributions fitting parameters
%
\begin{table*}[!]
\begin{minipage}[t]{\hsize}
\setlength{\extrarowheight}{3pt}
\caption{$n-$distribution fitting parameters}
\label{lines_table}
\centering
\small
\renewcommand{\footnoterule}{}
\begin{tabular}{l c c c c c c}
\hline \hline
$v$ (km s$^{-1}$) &  $c_{1}(v)$\tablefootmark{a}  &  $c_{2}(v)$\tablefootmark{a} & $c_{3}(v)$\tablefootmark{a} & $c_{4}(v)$\tablefootmark{a} \\
\hline
50   & -0.46 & 10.79 & -60.37 & -37.08 \\
80   & -0.37 & 9.38  & -58.85 & -18.09 \\
120  & -0.40 & 9.29  & -48.03 & -33.64 \\
200  & -0.31 & 7.73  & -44.56 & -21.31 \\
350  & -0.21 & 5.68  & -37.53 & -14.48 \\
600  & -0.17 & 3.39  & -30.87 & -0.82  \\
900  & -0.24 & 1.49  & -14.59 & -18.47 \\
1300 & -0.30 & 1.16  & -10.12 & -10.05 \\
1800 & -0.30 & 0.58  & -7.95  & -0.32  \\
2400 & -0.36 & 0.64  & -6.75  & 4.11   \\
3100 & -0.34 & 0.70  & -4.60  & -2.77  \\
3900 & -0.37 & 0.84  & -4.06  & -7.24  \\
4800 & -0.39 & 0.83  & -4.06  & -10.45 \\
5800 & -0.40 & 0.63  & -3.90  & -10.66 \\
\hline
\hline

\end{tabular}
\end{minipage}
 \tablefoot{
   \tablefoottext{a}{Parameters of the polynomial function (Eq.A.1).}
}
\end{table*}

%============================
%  FIG: Small-n
%
\begin{figure*}[!]
\centering
\resizebox{0.7\hsize}{!}{\hspace{-1cm}\includegraphics[angle=0]{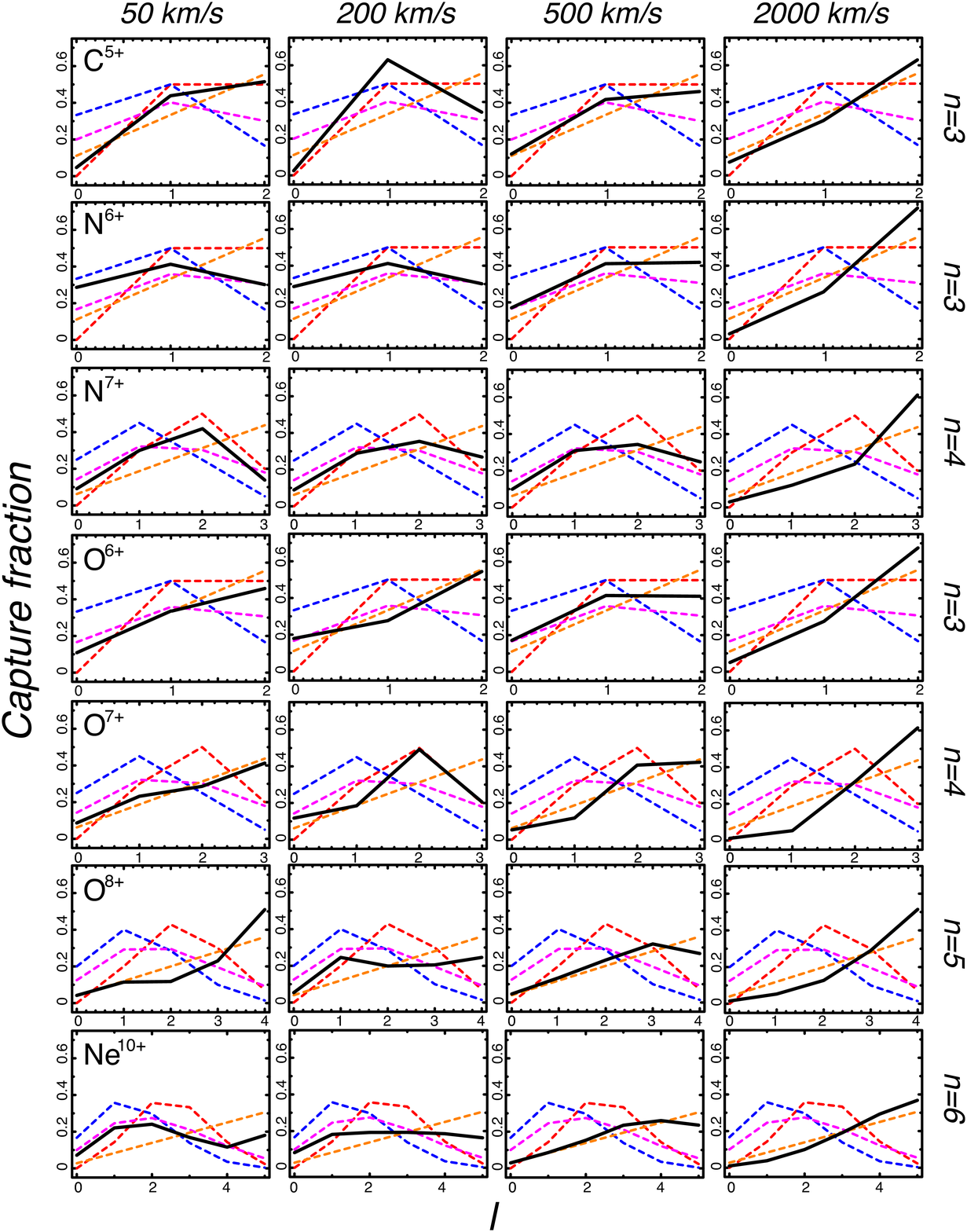}}
\caption{Averaged fractions of $l-$dependent capture for $\rm C^{5+}$, $\rm N^{6+}$, $\rm N^{7+}$, $\rm O^{6+}$, $\rm O^{7+}$, $\rm O^{8+}$, and $\rm Ne^{10+}$, plotted as a function of $l$. For each ion, the dominant $n-$shell in $n \leq n_{\rm p}$ is shown. The color and line styles are the same as used in Figure 3. 
\vspace{0.5cm}
}
\label{c2000_fig}
\end{figure*}
%============================

%============================
%  FIG: Big-n
%
\begin{figure*}[!]
\centering
\resizebox{0.7\hsize}{!}{\hspace{-1cm}\includegraphics[angle=0]{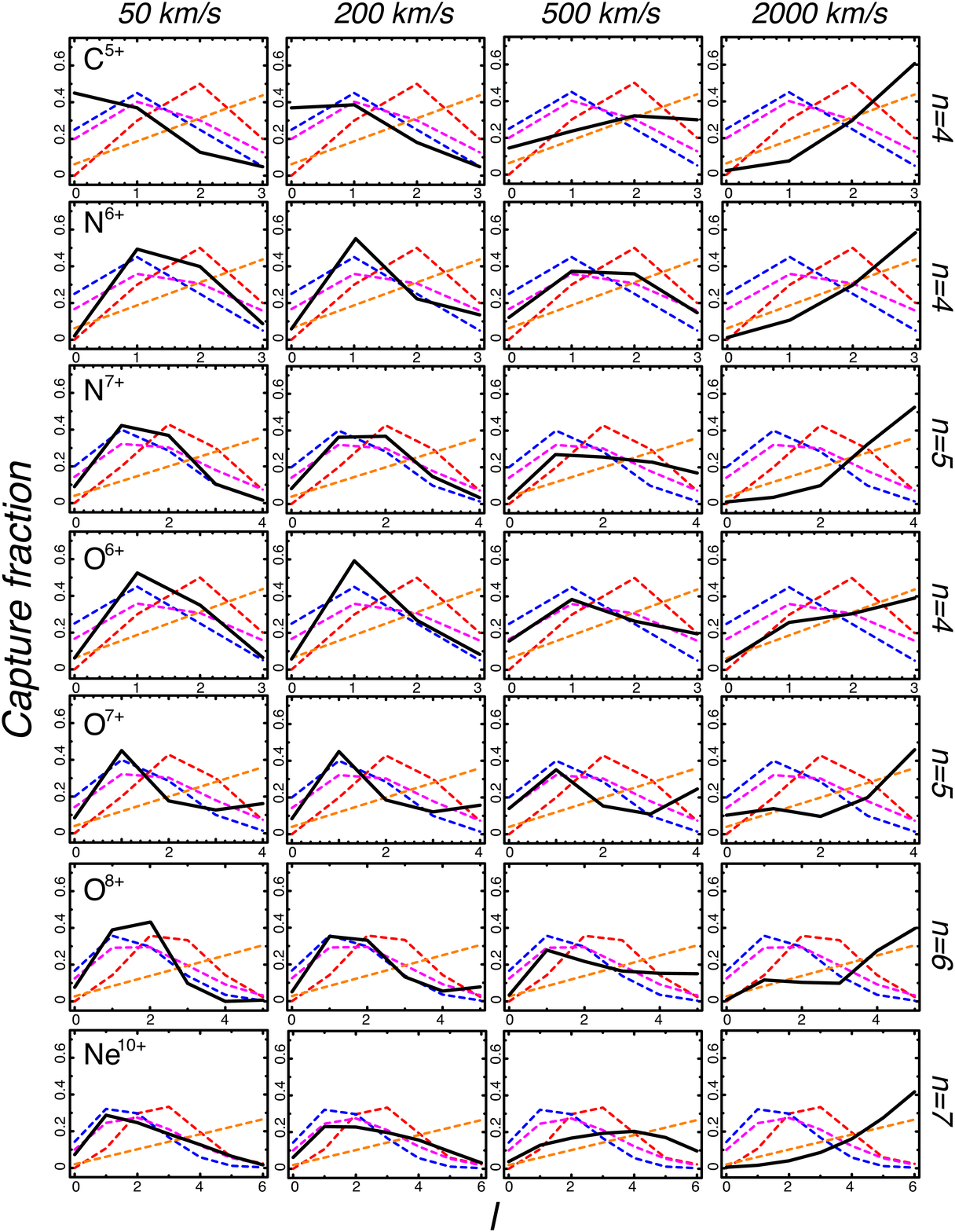}}
\caption{Averaged fractions of $l-$dependent capture for $\rm C^{5+}$, $\rm N^{6+}$, $\rm N^{7+}$, $\rm O^{6+}$, $\rm O^{7+}$, $\rm O^{8+}$, and $\rm Ne^{10+}$, plotted as a function of $l$. For each ion, the dominant $n-$shell in $n > n_{\rm p}$ is shown. The color and line styles are the same as used in Figure 3. 
\vspace{0.5cm}
}
\label{c2000_fig}
\end{figure*}
%============================

It has long been known that the CX cross section often distributes continuously as a function of $n$, with a maximum near $n_{\rm p}$, when
the ion charge $q$ is sufficiently large (see Janev \& Winter 1985 for a review). To quantify the $n$ distribution, 
we compiled in \S3.2 the velocity- and $n$- dependent cross sections found in literatures, and fit them with a phenomenological third-degree polynomial curve,
\begin{equation}
  \begin{aligned}
    \lg \sigma_{\rm norm}(n) = {} & c_{1}(v) + c_{2}(v) n_{\rm norm}(n,q)  + c_{3}(v) n_{\rm norm}^{2}(n,q) \\
                                & + c_{4}(v) n_{\rm norm}^{3}(n,q),
  \end{aligned}
\end{equation} 
where $\sigma_{\rm norm}(n)$ is the capture fraction into $n$, $c_{1}(v)$ to $c_{4}(v)$ are four velocity-dependent fitting parameters, and
$n_{\rm norm}(n,q) = (n - n_{\rm p})/n_{\rm p}$. The best-fit parameters are shown in Table A.1.

%====================%
\section{Preferred $l-$distributions}
%====================%

Here we describe in details a velocity-dependent scheme to approximate the $l-$ selectivity. As shown in Table 1, the
$l-$ resolved cross section data, derived from theoretical calculation, are available for reactions with the seven ions, i.e.,
$\rm C^{5+}$, $\rm N^{6+}$, $\rm N^{7+}$, $\rm O^{6+}$, $\rm O^{7+}$, $\rm O^{8+}$, and $\rm Ne^{10+}$. For other ions, the five
canonical weighting functions, as shown in Eqs.4$-$8, were utilized as follows.

For each velocity, we determined the preferred weighting function by
comparing them to the available data. As described in \S3.3, the preferred function must be chosen separately for shells with
principle quantum number $n \leq n_{\rm p}$ and those with $n > n_{\rm p}$. Here we only study the most dominant shell in each group (Figures A.1 and A.2).
For $n \leq n_{\rm p}$, the $W^{\rm se}$ function (Eq.6) is recommended
at low velocities, i.e., $v = 50$ and 200 km s$^{-1}$, while the statistical weight $W^{\rm st}$ becomes more popular in the intermediate- and high-
velocity regimes ($v = 500$ and 2000 km s$^{-1}$). As shown in Figure A.1, above scheme can be applied approximately to most reactions, except for a few
outliers such as $\rm C^{5+}$ and $\rm O^{8+}$ at $v = 50$ km s$^{-1}$, $\rm O^{6+}$ at $v = 200$ km s$^{-1}$, and $\rm N^{7+}$ at $v = 500$ km s$^{-1}$.
As for $n > n_{\rm p}$, the $l-$ distribution is best represented by $W^{\rm L2}$ at $v = 50$ and 200 km s$^{-1}$, $W^{\rm se}$ at $v = 500$ km s$^{-1}$,
and $W^{\rm st}$ at $v = 2000$ km s$^{-1}$, albeit with a few exceptions such as $\rm C^{5+}$ and $\rm Ne^{10+}$ at $v = 500$ km s$^{-1}$.

In order to define the $l-$ preference continuous in the velocity space, we have further analyzed the data with a finer velocity grid of 20 points.
It is found that the $l-$ distributions for $n \leq n_{\rm p}$ shells mostly switch at $v = 500$ km s$^{-1}$
from a $W^{\rm se}$ form to a $W^{\rm st}$ form, while the $n > n_{\rm p}$ shells are likely to evolve from $W^{\rm l2}$ to $W^{\rm se}$ at $v = 300$ km s$^{-1}$, and
from $W^{\rm se}$ to $W^{\rm st}$ at $v = 500$ km s$^{-1}$.

%====================%
\section{More features of the SPEX-CX model}
%====================%

Our plasma code for CX emission is included as an independent model in the SPEX package (Kaastra et al. 1996).\footnote{https://www.sron.nl/spex} 
The model first calculates the fraction of each ion at a ionization temperature $T_{\rm i}$ and element abundance $A$, then evaluates the CX spectrum 
for a collision velocity $v$ and emission measure $norm$. The rate coefficients obtained in \S3 are fully utilized in the model, and the actual data
are tabulated in forms of FITS files \footnote{http://fits.gsfc.nasa.gov/fits\_home.html} in the SPEX database. They will be updated 
with more recent published results once these become available.

The CX model contains three additional parameters for different physical conditions. First, the collision velocity $v$ can be replaced by 
the velocity of random thermal motion, which is characterized by ion temperature $T_{i}$. This is appropriate for some hot plasmas where thermal motion dominates.
Secondly, besides the single collision mode, our model also allows multiple collisions between ions and neutrals. In the latter case, one ion
would continuously undergo CX and produce various emission lines, until it becomes neutral. This is more suited for objects with dense neutral materials.    
Finally, the model provides five types of $l-$ weighting functions (Eqs.4-8) for the CX cross section. The optimized method is described in \S3.3, while
the five basic functions can also be selected to fine-tune the spectrum, and to test the sensitivity of data to the assumed subshell populations.

\end{document}